\documentclass[twocolumn]{aastex62}

\usepackage{amsmath}
\hypersetup{breaklinks=true,colorlinks=true,citecolor=blue,linkcolor=blue,urlcolor=blue}

\def\hst{\textit{HST}}

\shorttitle{The late-time lightcurves of SNe Ia}
\shortauthors{Graur et al.}

\begin{document}

\title{Observations of SN 2015F suggest a correlation between the intrinsic luminosity of Type I\MakeLowercase{a} supernovae and the shape of their light curves $>900$ days after explosion}

\correspondingauthor{Or Graur}
\email{or.graur@cfa.harvard.edu}

\author{Or Graur}
\affiliation{Harvard-Smithsonian Center for Astrophysics, 60 Garden St., Cambridge, MA 02138, USA}
\affiliation{Department of Astrophysics, American Museum of Natural History, New York, NY 10024, USA}

\author{David R. Zurek}
\affiliation{Department of Astrophysics, American Museum of Natural History, New York, NY 10024, USA}

\author{Armin Rest}
\affiliation{Space Telescope Science Institute, Baltimore, MD 21218, USA}

\author{Ivo R. Seitenzahl}
\affiliation{School of Physical, Environmental, and Mathematical Sciences, University of New South Wales, Australian Defense Force Academy, Canberra, ACT 2600, Australia}
\affiliation{Research School of Astronomy and Astrophysics, The Australian National University, Cotter Road, Weston Creek, ACT 2611, Australia}

\author{Benjamin J. Shappee}
\altaffiliation{Hubble and Carnegie-Princeton Fellow}
\affiliation{The Observatories of the Carnegie Institution for Science, 813 Santa Barbara St., Pasadena, CA 91101, USA}
\affiliation{Institute for Astronomy, University of Hawai'i, 2680 Woodlawn Drive, Honolulu, HI 96822, USA}

\author{Robert Fisher}
\affiliation{Department of Physics, University of Massachusetts Dartmouth, 285 Old Westport Road, North Dartmouth, MA 02740, USA}
\affiliation{Harvard-Smithsonian Center for Astrophysics, 60 Garden St., Cambridge, MA 02138, USA}

\author{James Guillochon}
\affiliation{Harvard-Smithsonian Center for Astrophysics, 60 Garden St., Cambridge, MA 02138, USA}

\author{Michael M. Shara}
\affiliation{Department of Astrophysics, American Museum of Natural History, New York, NY 10024, USA}

\author{Adam G. Riess}
\affiliation{Space Telescope Science Institute, Baltimore, MD 21218, USA}
\affiliation{Department of Physics and Astronomy, The Johns Hopkins University, Baltimore, MD 21218, USA}


\begin{abstract}
The late-time light curves of Type Ia supernovae (SNe Ia), observed $>900$ days after explosion, present the possibility of a new diagnostic for SN Ia progenitor and explosion models. First, however, we must discover what physical process (or processes) leads to the slow-down of the light curve relative to a pure $^{56}$Co decay, as observed in SNe 2011fe, 2012cg, and 2014J. We present \textit{Hubble Space Telescope} observations of SN 2015F, taken $\approx600$--$1040$ days past maximum light. Unlike those of the three other SNe Ia, the light curve of SN 2015F remains consistent with being powered solely by the radioactive decay of $^{56}$Co. We fit the light curves of these four SNe Ia in a consistent manner and measure possible correlations between the light curve stretch---a proxy for the intrinsic luminosity of the SN---and the parameters of the physical model used in the fit. We propose a new, late-time Phillips-like correlation between the stretch of the SNe and the shape of their late-time light curves, which we parametrize as the difference between their pseudo-bolometric luminosities at $600$ and $900$ days: $\Delta L_{900} = {\rm log}(L_{600}/L_{900})$. Our analysis is based on only four SNe, so a larger sample is required to test the validity of this correlation. If ture, this model-independent correlation provides a new way to test which physical process lies behind the slow-down of SN Ia light curves $>900$ days after explosion, and, ultimately, fresh constraints on the various SN Ia progenitor and explosion models. 
\end{abstract}



\section{Introduction}
\label{sec:intro}

There is ample evidence that Type Ia supernovae (SNe Ia) are thermonuclear explosions of carbon-oxygen white dwarfs (WDs): see, e.g., the review by \citet{2014ARA&A..52..107M}. However, the question remains how, exactly, the WD is disrupted. Over the years, various progenitor and explosion scenarios have been suggested, which gave rise to a host of observational diagnostics. These include, among others, reconstructions of the SN Ia delay-time distribution (e.g., \citealt{2017ApJ...848...25M} and references therein), pre-explosion imaging (e.g., \citealt{Li2011fe,2014MNRAS.442L..28G,2014ApJ...790....3K}), and strong constraints from radio and X-ray observations (e.g., \citealt{2012ApJ...746...21H,2012ApJ...751..134M,2014ApJ...790...52M,2016ApJ...821..119C}). In the last few years, a new diagnostic tool has begun to emerge: the late-time light curves of SNe Ia, as measured $>900$ days after explosion.

The light we receive from SNe Ia is predominantly produced by $\gamma$ rays emitted by the radioactive decay chain $^{56}$Ni$\to ^{56}$Co$\to ^{56}$Fe, which thermalize in the SN ejecta and get re-emitted in the optical and near-infrared (NIR). As the ejecta expand, they become optically thin to the high-energy photons, and heating proceeds via positron trapping \citep{1979ApJ...230L..37A,1993ApJ...405..614C,1997A&A...328..203C,1999ApJS..124..503M}. 

\citet{1980PhDT.........1A} predicted that, around $500$ days past maximum light, the cooling of the SN ejecta, which until this time proceeded through transitions in the optical, should switch to fine-structure transitions in the IR, resulting in a steep drop in the optical light curve. Tantalizingly, the very last data points on the light curves of SN 1992A \citep{1997A&A...328..203C} and SN 2003hv \citep{2009A&A...505..265L} indicated that this ``IR catastrophe'' might not actually take place.

Instead, \citet{2009MNRAS.400..531S} suggested that $>900$ days past explosion, SN Ia light curves should be boosted by thermalization of X-rays, as well as internal-conversion and Auger electrons, emitted by the long-lived decay chains $^{57}{\rm Co}\to^{57}{\rm Fe}$ ($t_{1/2}\approx272$ days) and $^{55}{\rm Fe}\to^{55}{\rm Mn}$ ($t_{1/2}\approx1000$ days). The density of the WD prior to explosion will affect at what ratios, relative to $^{56}{\rm Ni}$, these isotopes are produced during the explosion. \citet{2012ApJ...750L..19R} used this assumption to predict what the late-time light curve of SN 2011fe would look like in the case of either the ``single-degenerate'' (a WD accreting matter from a non-degenerate companion; \citealt{Whelan1973,1985ApJ...297..531N}) or ``double-degenerate'' (in which two WDs merge or collide due to loss of angular momentum and energy to gravitational waves; \citealt{Iben1984,Webbink1984}. \citealt{2012ApJ...750L..19R} specifically tested a merger of two carbon-oxygen WDs) progenitor scenarios. 

\citet{2016ApJ...819...31G} followed the late-time light curve of SN 2012cg out to $1055$ days and showed that it slowed down exactly as \citet{2009MNRAS.400..531S} had predicted. Similar observations were also carried out for SN 2011fe \citep{2014ApJ...796L..26K,2017MNRAS.472.2534K,2017MNRAS.468.3798D,2017ApJ...841...48S} and SN 2014J \citep{2018ApJ...852...89Y}. In all three of these SNe, the model suggested by \citet{2009MNRAS.400..531S} was consistent with the observations. Furthermore, \citet{2017ApJ...841...48S} measured late-time photometry of SN 2011fe that was precise enough to rule against the N100 delayed-detonation model (a single-degenerate scenario) computed by \citet{2012ApJ...750L..19R}.

And yet, the slow-down of the late-time light curves of these SNe can also be explained by other means. Light echoes could contaminate the light curves and cause them to flatten out starting at $\sim 500$ days (e.g., \citealt{2005MNRAS.357.1161P}). Such contamination was convincingly ruled out for SN 2011fe (e.g., \citealt{2017ApJ...841...48S}). \citet{2015ApJ...804L..37C} discovered a resolved light echo from SN 2014J, but \citet{2018ApJ...852...89Y} argued that it does not contaminate the unresolved light from the SN. \citet{2016ApJ...819...31G} could not rule out a light echo for SN 2012cg; however, in this work we show that it too was not contaminated.

\citet{1993ApJ...408L..25F} suggested that at late times, the low density of the ejecta would lead to recombination and cooling timescales longer than the radioactive decay timescale, and the conversion of energy absorbed by the ejecta into emitted radiation would no longer be instantaneous. \citet{2015ApJ...814L...2F} showed that taking this ``freeze-out'' effect into account allowed them to model a late-time spectrum of SN 2011fe observed by \citet{2015MNRAS.448L..48T} $1034$ days after maximum light. \citet{2017MNRAS.472.2534K} showed that freeze-out, combined with $^{56}$Co decay, was also consistent with the light curve of SN 2011fe, thus providing an alternative explanation to the $^{57}$Co-powered tail.

Finally, it is still unclear what fraction of positrons is trapped by the ejecta as it continues to expand at late times. Both \citet{2017MNRAS.472.2534K} and \citet{2017MNRAS.468.3798D} showed that varying the fraction of positron trapping could lead to light curve models consistent with the observations of SN 2011fe. 

The observations of SNe 2011fe, 2012cg, and 2014J have established a strong case against the occurrence of the ``IR catastrophe.'' However, it remains to be seen what physical process (or processes) lies behind the slow-down of the light curve $>900$ days after explosion, and how late-time light curves could be used to constrain the multitude of SN Ia progenitor and explosion models.

In this work, we present \textit{Hubble Space Telescope} (\hst) observations of SN 2015F between $\sim600$ and $\sim1040$ days past maximum light. In Section~\ref{sec:sample}, we show that, as opposed to previous SNe Ia, the pseudo-bolometric light curve of SN 2015F is consistent with being driven solely by the radioactive decay of $^{56}{\rm Co}\to^{56}$Fe. This leads us to study the light curves of SNe 2011fe, 2012cg, 2014J, and 2015F as a collective group. 

In Section~\ref{sec:decay}, we show that whatever physical process is invoked to fit the late-time light curves, there appears to be a correlation between the main parameter of that model (e.g., the mass ratio between $^{56}$Co and $^{57}$Co or the time at which freeze-out sets in) and the stretch of the SN, a proxy for its intrinsic luminosity. Inspired by the famous ``width-luminosity'' correlation \citep{1993ApJ...413L.105P}, we propose a similar correlation between the stretch of a SN Ia and the shape of its late-time light curve, with the shape parametrized by the difference between the pseudo-bolometric luminosities at $600$ and $900$ days past maximum light: $\Delta L_{900} = {\rm log}(L_{600}/L_{900})$.\footnote{See also \citet{2013ApJ...767..119M}, who found a correlation between $\Delta m_{15}(B)$ and the color index of Spitzer mid-infrared light curves of four SNe Ia observed roughly a year after explosion.} In Section~\ref{sec:discuss}, we conclude that, if corroborated by observations of more SNe Ia, the correlations presented here (especially the model-independent $\Delta L_{900}$ vs.\ stretch correlation) could be used as a new diagnostic for SN Ia nebular-phase physics, as well as progenitor and explosion models.

\begin{figure*}
 \centering
 \includegraphics[width=0.94\textwidth]{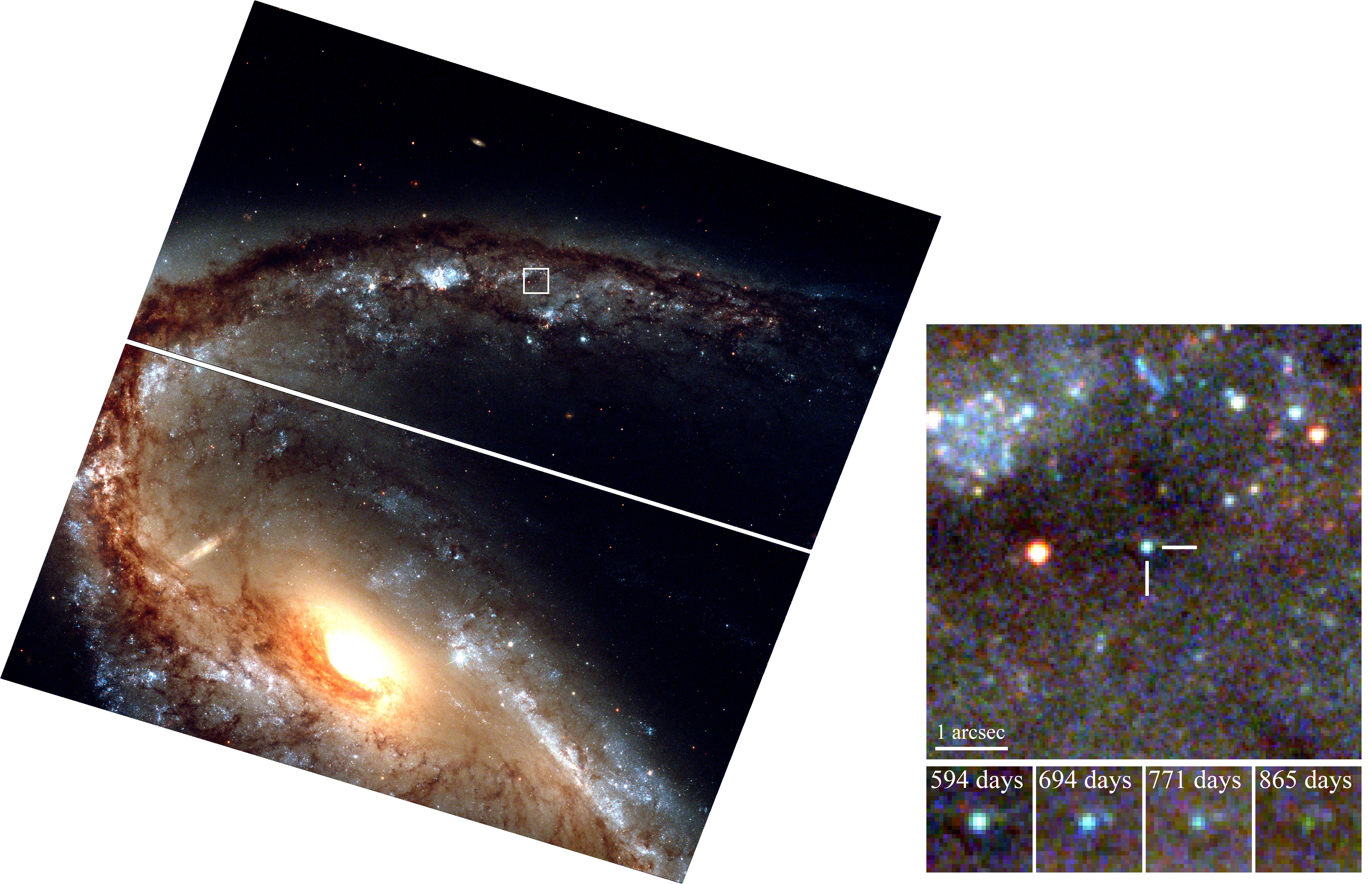}
 \caption{Left panel: \textit{HST} color composite of SN 2015F in the northern half of NGC 2442. The image is composed of $F814W$ (red), $F555W$ (green), and $F438W$ (blue) images from the first visit to NGC 2442 in program GO--14611. The location of the SN in the spiral arm of the galaxy is marked by a white box. Top right panel: the $6^{\prime\prime}\times6^{\prime\prime}$ region hemmed in by the white box in the left panel. SN 2015F, identified by the white reticle, exploded in a relatively isolated region of the spiral arm. Bottom right panel: successive visits show SN 2015F fading over $270$ days. On the last visit shown here, the SN is no longer detected in both $F438W$ and $F814W$. Each tile is $1^{\prime\prime}$ on a side. In all the panels shown here, North is up and East is to the left.}
 \label{fig:host}
\end{figure*}

\begin{figure*}
 \centering
 \begin{tabular}{cc}
  \includegraphics[width=0.47\textwidth]{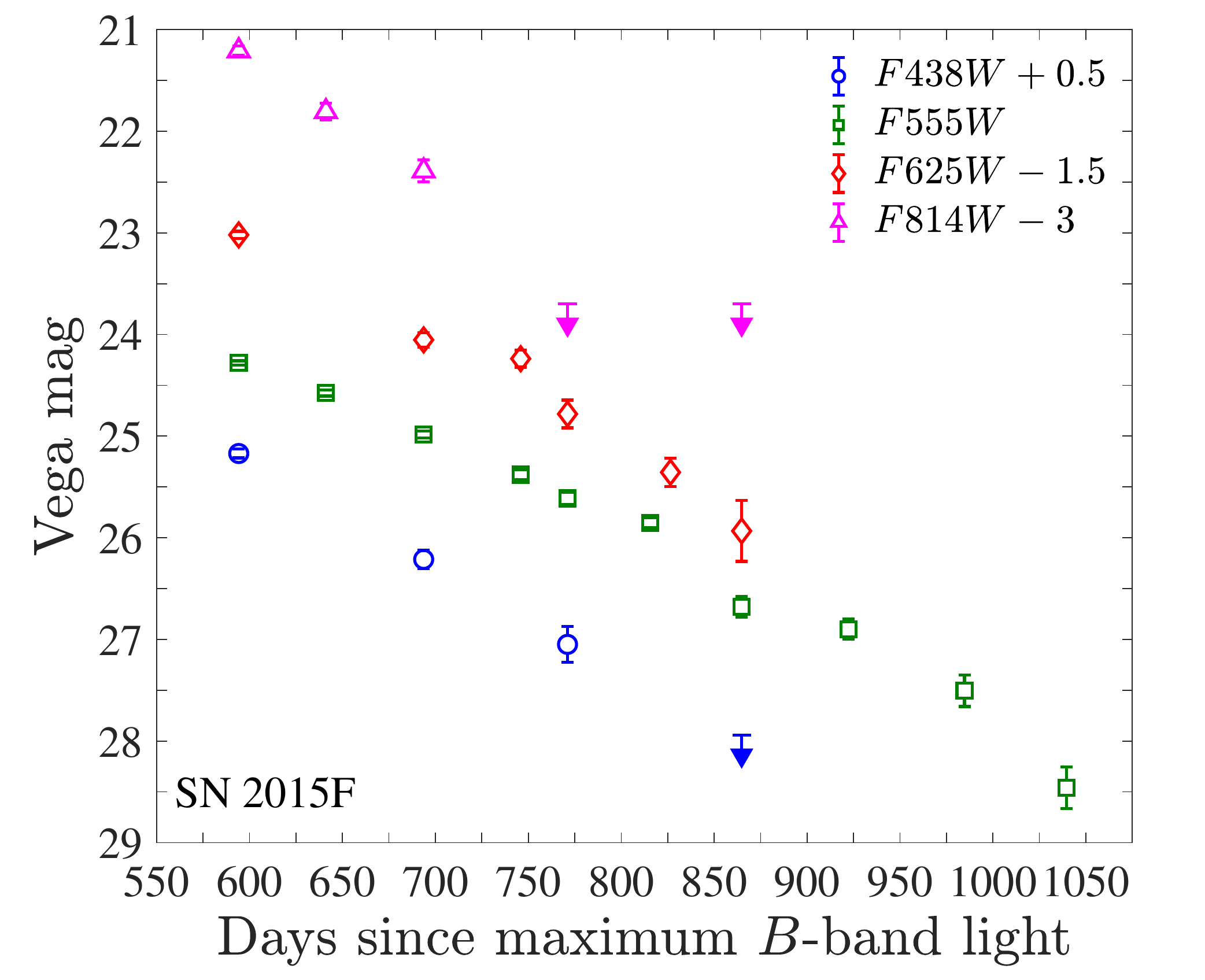} & \includegraphics[width=0.47\textwidth]{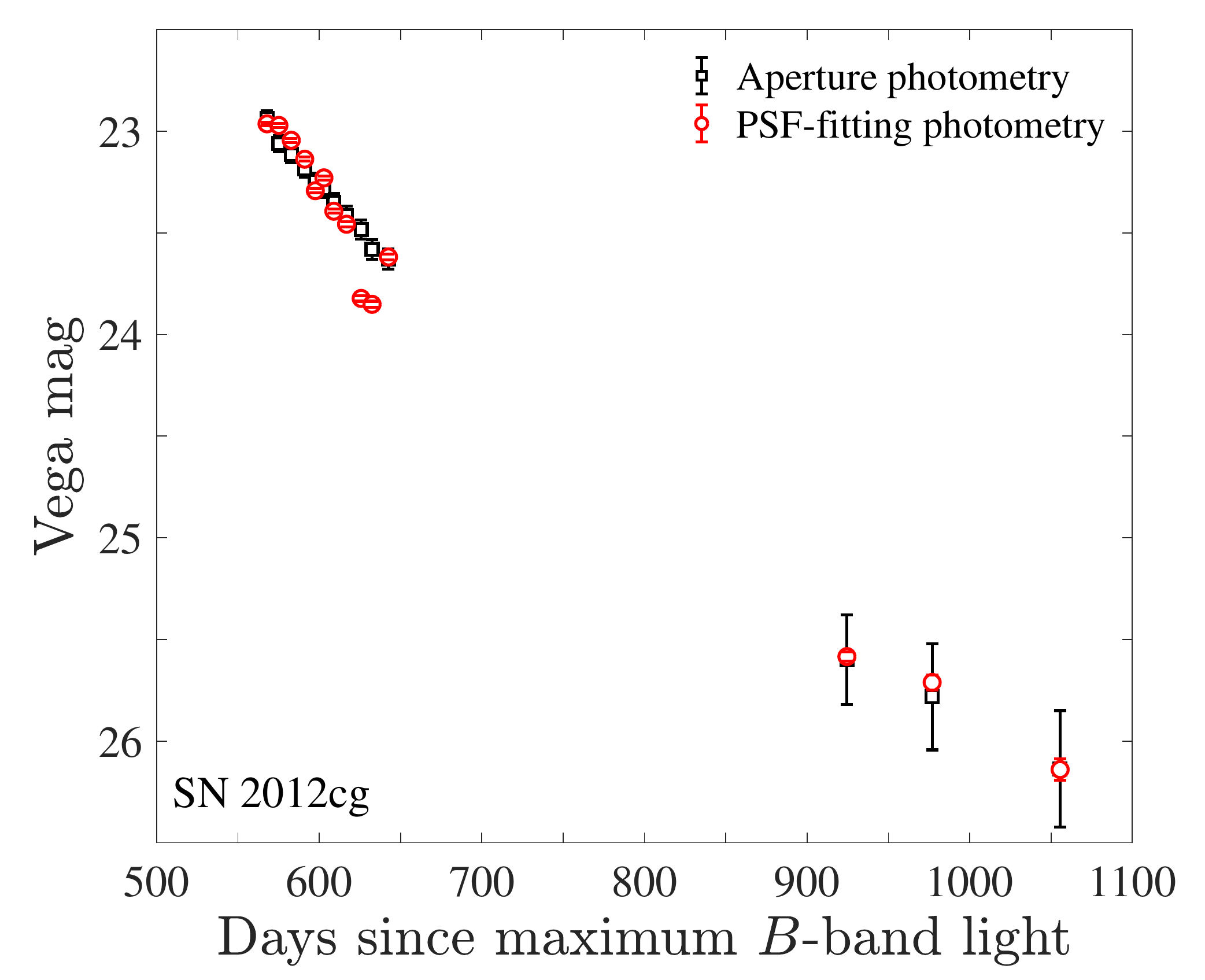} \\
 \end{tabular}
 \caption{Left panel: PSF-fitting photometry of SN 2015F in $F438W$ (blue circles), $F555W$ (green squares), $F625W$ (red diamonds), and $F814W$ (magenta triangles). Downturned arrows are upper limits defined as the magnitude at which a point source would have a S/N ratio of 3. Right panel: PSF-fitting photometry (red circles) of SN 2012cg, compared to the aperture photometry (white squares) from \citet{2016ApJ...819...31G}. The PSF-fitting photometry has been scaled up by $0.18$ mag to fit the aperture photometry at $600$ days. {\sc Dolphot} masked some of the SN pixels in the {\sc flt} images on $625.8$ and $632.5$ days, leading to systematically low fluxes. The measurements in both panels have not been corrected for Galactic or host-galaxy extinction.}
 \label{fig:mag}
\end{figure*}


\section{Observations and Photometry}
\label{sec:sample}

SN 2015F was discovered on 2015 March 9 (UT) in the nearby spiral galaxy NGC 2442 by \citet{2015CBET.4081....1M} and classified as a SN Ia by \citet{2015ATel.7209....1F}. The host galaxy has several measured distances. The NASA Extragalactic Catalog \citep{1991ASSL..171...89H} lists a Tully-Fisher distance of $17.1$ Mpc. Based on the light curve of the SN, \citet{2015ApJS..221...22I} and  \citet{2017MNRAS.464.4476C} measured $23.9\pm0.4$ Mpc and $21$--$22$ Mpc, respectively. Here, we use a mean distance of $20.4$ Mpc. \citet{2015ApJS..221...22I} and \citet{2017MNRAS.464.4476C} measured consistent $\Delta m_{15}(B)$ values of $1.26 \pm 0.10$ and $1.35 \pm 0.03$ mag, respectively, placing SN 2015F between ``normal'' and subluminous SN 1991bg-like SNe Ia. A nebular spectrum taken $279$ days past maximum light confirms that SN 2015F is a normal, though subluminous, SN Ia \citep{2017MNRAS.472.3437G}. 

We imaged SN 2015F ($\alpha=07^{\rm h}36^{\rm m}15.76^{\rm s}$, $\delta=-69^{\circ}30^{\prime}23.0^{\prime\prime}$) with the \hst\ Wide Field Camera 3 (WFC3) wide-band filters {\it F438W}, {\it F555W}, {\it F625W}, and {\it F814W} under \hst\ programs GO--14611 and GO--15415 (PI Graur) on eleven separate occasions between 2016 November 8 and 2018 January 28. At these times, the SN was $594$--$1040$ days past maximum light. A complete list of dates and phases of the observations is presented in Table~\ref{table:mags_15F}. These observations can be found in the Mikulski Archive for Space Telescopes (MAST) at \href{http://archive.stsci.edu/doi/resolve/resolve.html?doi=10.17909/T9109B}{DOI:10.17909/T9109B}.

In Figure~\ref{fig:host}, we show a color composite of NGC 2442 and SN 2015F along with tiles that show the SN fading between $594$ and $865$ days. Throughout this work, the phases we cite are calculated relative to $B$-band maximum light, which occurred on 2015 March 25 (JD 57106.5; \citealt{2015ApJS..221...22I}).

\subsection{PSF-fitting photometry}
\label{subsec:phot}

Using the {\sc tweakreg} and {\sc astrodrizzle} routines included in the {\sc drizzlepac} pyraf package,\footnote{\url{http://drizzlepac.stsci.edu/}} we aligned the \hst\ images and removed cosmic rays and bad pixels. Next, we performed point-spread-function (PSF) fitting photometry of SN 2015F using {\sc Dolphot}\footnote{\url{http://americano.dolphinsim.com/dolphot/}} \citep{2000PASP..112.1383D} and the {\sc flc} files produced by the \hst\ WFC3 pipeline, which are corrected for charge transfer efficiency effects. In each filter, we first ran {\sc Dolphot} on the images from the first visit, in which the SN was brightest and so easiest to center. We then forced {\sc Dolphot} to photometer the SN at the same fixed location. At 864.8 and 1039.5 days, {\sc Dolphot} failed to detect the SN in its forced location. Allowing {\sc Dolphot} to center the PSF of the SN on its own results in centroids only $\sim 0.2$ and $\sim 0.6$ pixels away from the forced centroid in \textit{F555W} and \textit{F625W}, respectively. The resulting photometry, in Vega mags, is presented in Table~\ref{table:mags_15F} and Figure~\ref{fig:mag}.

We repeat these steps for the $F350LP$ data of SN 2012cg, in order to reduce the statistical uncertainties of its photometry. However, we use the original {\sc flt} instead of {\sc flc} files, as the images at $>900$ days did not have the latter type of file on the Mikulski Archive for Space Telescopes. Because the $F350LP$ data before and after 900 days were taken by two different programs (GO--12880 and GO--13799, respectively), which used different UVIS apertures, we test for systematic differences between the datasets. {\sc Dolphot} identifies 4863 point sources within a $500\times500$ pixel$^2$ area centered on SN 2012cg. For each of these objects, we measure the median of its $F350LP$ photometry in each visit, pre- and post-$900$ days, and find that the $>900$-day photometry is systematically brighter by $\sim 0.1$ mag. We correct the $>900$-day photometry accordingly. The resultant photometry is presented in Table~\ref{table:mags_12cg} and in Figure~\ref{fig:mag}.

\citet{2014ApJS..215....9W} and \citet{2017ApJ...841...48S} noted that {\sc Dolphot} tends to underestimate uncertainties by factors of a few (though see \citealt{2017MNRAS.472.2534K}, who, unlike \citealt{2017ApJ...841...48S}, do not measure this effect for SN 2011fe). We followed the method outlined by \citet{2017ApJ...841...48S} to test for this effect, and found that the uncertainties on the photometry of SN 2015F were estimated correctly. A similar test of the SN 2012cg photometry provided the same result. 

\citet{2017ApJ...841...48S} accounted for crowding of SN 2011fe by background stars, especially in the NIR. We do not test for such an effect here, for two reasons: 1) a visual inspection shows that the SN is relatively isolated from other stars (Figure~\ref{fig:host}); and 2) such an effect, if present, would result in the addition of a constant flux to our photometry. In this work, we only diagnose the \emph{shape} of the light curve, and so any systematic offset will have no effect on our results.

\subsection{Pseudo-bolometric light curve}
\label{subsec:pbl}

We follow \citet{2017ApJ...841...48S} and construct a pseudo-bolometric light curve from our \textit{HST} observations. Because we only imaged SN 2015F in all four filters every other visit, we begin by linearly interpolating (or, after the SN is no longer visible in a specific filter, linearly extrapolating) the missing photometry in the intervening visits. 
In order to extrapolate the missing observations, we make two assumptions:
\begin{enumerate}
 \item Similar to SN 2011fe, SN 2015F has a flat color evolution in the time period over which we extrapolate. This assumption is supported, in part, by a comparison of the $V-R$ colors of the two SNe, as shown in Section~\ref{subsec:le} below.
 \item The optical light from SN 2015F is dominated by the light in the \textit{F555W} band, corresponding to the [Fe II] and [Fe III] emission-line complex at $\sim 5300$ \AA~\citep{2015MNRAS.454.1948G}. This has been shown to be true for SNe 2011fe \citep{2014ApJ...796L..26K,2017ApJ...841...48S} and 2012cg \citep{2016ApJ...819...31G} at least out to $\sim 1100$ days.
\end{enumerate}
  
The measured and extrapolated magnitudes are corrected for Galactic extinction along the line of sight to SN 2015F and for host-galaxy extinction. The line-of-sight extinctions in $F438W$, $F555W$, $F625W$, and $F814W$ are $0.736$, $0.580$, $0.459$, and $0.312$ mag, respectively \citep{2011ApJ...737..103S}. Using the \citet{1989ApJ...345..245C} reddening law, a measured $E(B-V)=0.085 \pm 0.019$ mag \citep{2017MNRAS.464.4476C}, and assuming $R_V=3.1$, we estimate host-galaxy extinctions of $0.359$, $0.269$, $0.230$, and $0.157$ mag in the same filters.

Next, because there are no published spectra of SN 2015F at the phases probed here, we use the spectrum of SN 2011fe at 593 days to match our measured photometry. This spectrum, measured by \citet{2015MNRAS.454.1948G}, does not cover the full range of the \textit{F438W} filter. As \citet{2015MNRAS.454.1948G} showed, there is very little evolution in the spectrum of SN 2011fe between 593 and 981 days, so we use the blue part of the second spectrum to extend the first spectrum down to $3500$~\AA. Likewise, we extend the flat continuum at the end of the 593-day spectrum out to $10000$~\AA, in order to account for the width of the \textit{F814W} filter.

\begin{figure*}
 \centering
 \begin{tabular}{cc}
  \includegraphics[width=0.47\textwidth]{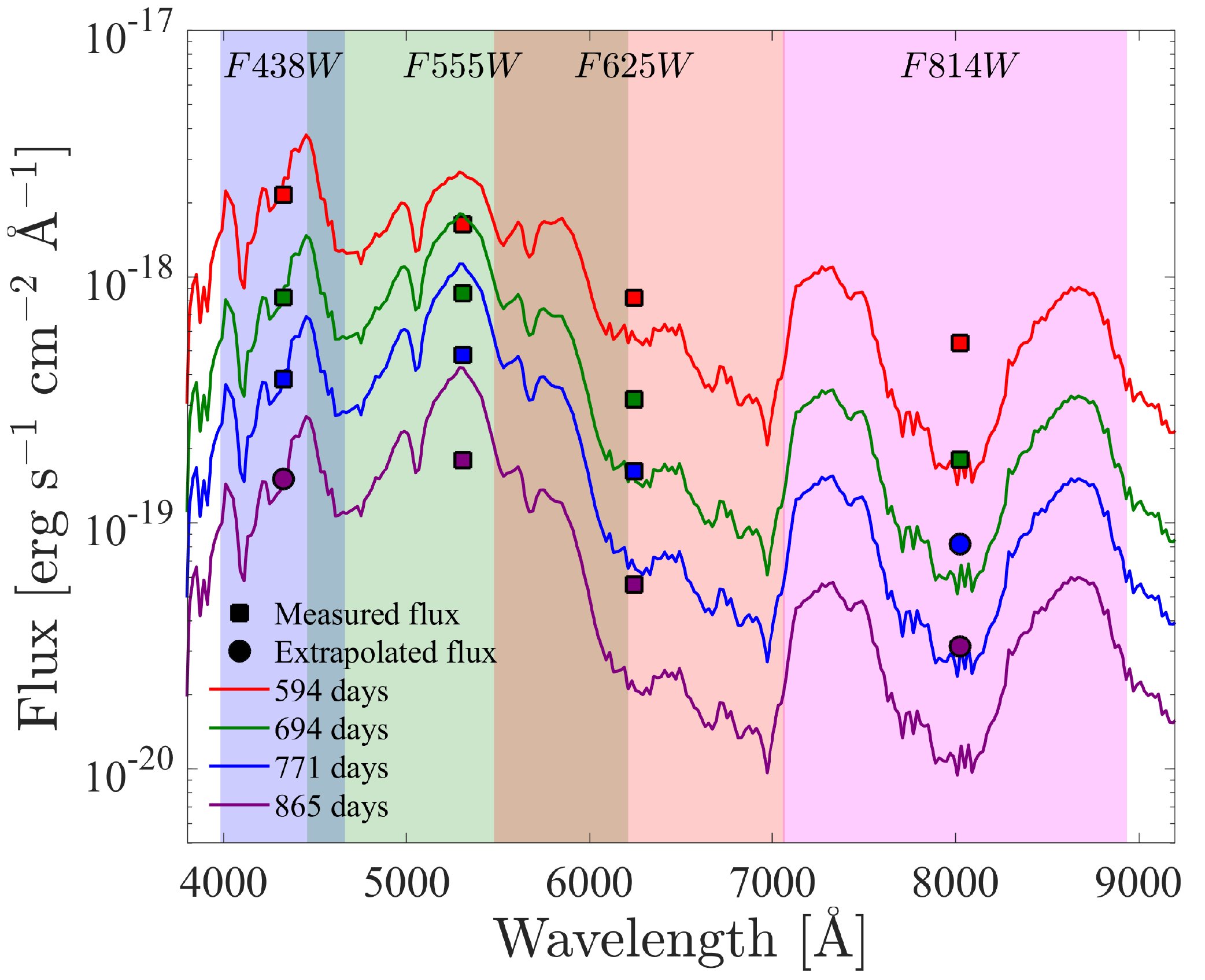} & \includegraphics[width=0.47\textwidth]{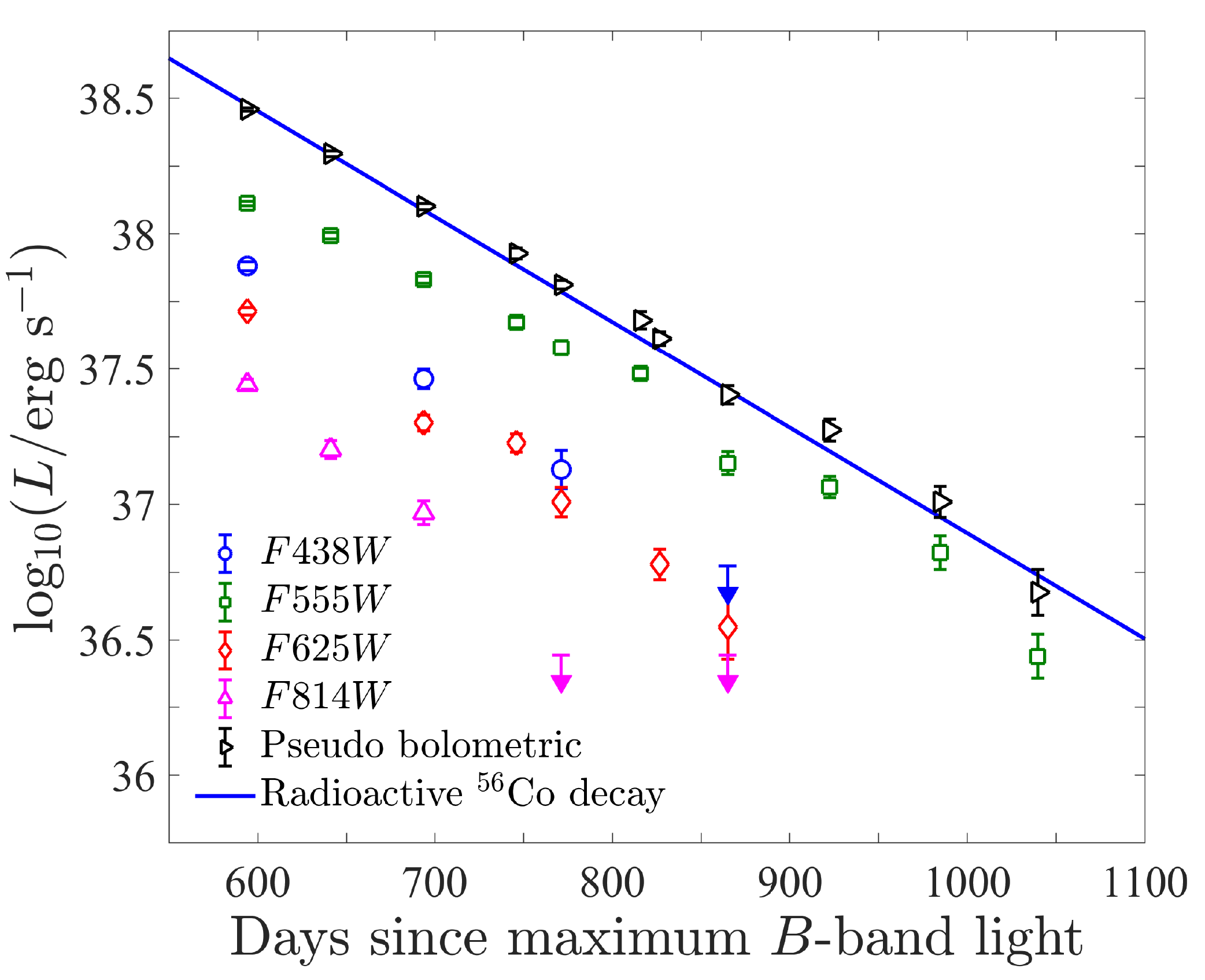} \\
 \end{tabular}
 \caption{Left panel: composite spectrum of SN 2011fe at 593 days morphed to fit the photometry of SN 2015F on four separate visits (solid curves). For display purposes, the spectra have been rebinned using a $20^{\prime\prime}$-long bin. Measured and extrapolated fluxes in each visit are shown as squares or circles, respectively. The spectrum is not required to pass through the flux measurements. Instead, synthetic photometry of the spectrum should yield results within $0.001$ mag of the measured fluxes. The colored bands represent the wavelength ranges where the filters used here have a throughput of $\geqslant 10$\%. Right panel: SN 2015F photometry converted to luminosities (symbols as in Figure~\ref{fig:mag}). The light curve is dominated by the $F555W$ luminosity. The resultant pseudo-bolometric light curve of SN 2015F (white right-facing triangles) is consistent with the pure radioactive decay of $^{56}$Co (blue solid curve). The $68$\% uncertainty band around the best-fitting curve is too thin to make out here.}
 \label{fig:pbl}
\end{figure*}

In each epoch, we morph the composite spectrum to fit the observed photometry. We require that the differences between the observed and synthetically measured photometry, $\Delta m$, be $<0.001$ mag at the pivot wavelengths of the \hst\ filters. In-between these wavelengths, we linearly interpolate the $\Delta m$ values to cover the entire spectrum. At $\lambda<4325$~\AA\ and $\lambda>8024$~\AA\ (the pivot wavelengths of \textit{F438W} and \textit{F814W}), we apply to the composite spectrum the $\Delta m$ values measured at those wavelengths. We show several examples of the resultant spectra in Figure~\ref{fig:pbl}.

Finally, we integrate the composite spectrum between $3500$--$10000$~\AA\ to derive the pseudo-bolometric magnitude in each visit. To estimate uncertainties on these magnitudes, we repeat the steps outlined above in a Monte Carlo simulation in which we vary the observed magnitudes in all four filters according to their uncertainties. The pseudo-bolometric luminosities are shown in Table~\ref{table:mags_15F} and in Figure~\ref{fig:pbl}. Where we extrapolated missing observations, we also extrapolated their uncertainties, so that the S/N ratio of the extrapolated observations decreases with time. These extrapolated uncertainties are then folded into the Monte Carlo simulation. Thus, if our assumptions about the color evolution and dominance of the light coming out in $F555W$ are correct, the final uncertainties of the pseudo-bolometric luminosities should be conservative overestimates.

The pseudo-bolometric light curve of SN 2015F is consistent with the pure radioactive decay of $^{56}$Co at least out to $1040$ days ($\chi^2/{\rm DOF}=18/10$). Of the four SNe Ia observed out to these late times, SN 2015F is the first for which this null hypothesis is not rejected outright.

\section{SN I\MakeLowercase{a} progenitor constraints}
\label{sec:decay}

In this section, we use the pseudo-bolometric light curve derived in Section~\ref{subsec:pbl}, along with similar light curves of SNe 2011fe, 2012cg, and 2014J, as a diagnostic of SN Ia nebular-phase physics. As described in Section~\ref{sec:intro}, the light curves of SNe Ia can begin to deviate from the radioactive decay of $^{56}$Co for several reasons. In Section~\ref{subsec:le}, we show that SNe 2015F and 2012cg are not contaminated by light echoes. Next, we show that, whether one assumes that the late-time light curve is boosted by radioactive decay of other isotopes of Ni (Section~\ref{subsec:decay}) or by freeze-out (Section~\ref{subsec:freeze}), there is a possible correlation between the results of these models for the SNe Ia used here and their intrinsic luminosity. Based on this observation, in Section~\ref{subsec:nonparam} we offer a model-independent correlation between the intrinsic luminosity of SNe Ia and the shape of their late-time light curves.

\subsection{Light echoes}
\label{subsec:le}

As mentioned in Section~\ref{sec:intro}, the most significant source of contamination of late-time SN Ia light curves is the possible presence of a light echo. Produced by the light of the SN reflected off dust sheets in the vicinity of the explosion, light echoes have been found to contaminate several SN Ia light curves $>500$ days after explosion (e.g., SNe 1991T, 1995E, 1998bu, 2006X, 2007af, and 2014J; \citealt{1994ApJ...434L..19S,1999ApJ...523..585S,2006ApJ...652..512Q,2001ApJ...549L.215C,2008ApJ...677.1060W,2015ApJ...805...71D,2015ApJ...804L..37C}). Once the light echo becomes the dominant source of light from the SN, it will cause the SN to appear bluer than it should at late times, and the bolometric light curve will flatten out \citep{2005MNRAS.357.1161P,2006MNRAS.369.1949P,2012PASA...29..466R}. 

SN 2011fe exploded in a very clean environment \citep{2012ApJ...750..164C}. \citet{2017ApJ...841...48S} further showed that the late-time colors of the SN were inconsistent with a light echo. \citet{2016ApJ...819...31G} showed that a combination of $^{56}$Co decay and a faint light echo that declined as $t^{-1}$ provided a similar fit to their measurements of SN 2012cg as the combination of $^{56}$Co and $^{57}$Co decays. Although SN 2014J is known to have a resolved light echo, \citet{2018ApJ...852...89Y} argue that the unresolved light of the SN is uncontaminated, based on a comparison of its colors with those of SN 2011fe (Y. Yang, private communication). Below, we show that light echoes can also be ruled out for both SNe 2015F and 2012cg.

\subsubsection{SN 2015F}
\label{susubbsec:le15F}

\floattable
\begin{deluxetable}{lCCcCCC}
 \tablecaption{Observation log for SN 2015F. \label{table:mags_15F}}
 \tablehead{
 \colhead{Date} & \colhead{MJD}    & \colhead{Phase}  & \colhead{Filter} & \colhead{Exposure Time} & \colhead{Magnitude}  & \colhead{Luminosity} \\
 \colhead{}     & \colhead{(days)} & \colhead{(days)} & \colhead{}       & \colhead{(s)}           & \colhead{(Vega mag)} & \colhead{${\rm log}(L/{\rm erg~s^{-1}})$} 
 }
 \decimals
 \startdata
  2016 Nov. 08.7 & 57700.7 & 594.2 & {\it F438W} & 1266 & 24.67 \pm 0.04 & 37.88 \pm 0.02 \\
  2016 Nov. 08.8 & 57700.8 & 594.3 & {\it F555W} & 1395 & 24.28 \pm 0.02 & 38.11 \pm 0.01 \\
  2016 Nov. 08.8 & 57700.8 & 594.3 & {\it F625W} & 1266 & 24.52 \pm 0.04 & 37.71 \pm 0.01 \\ 
  2016 Nov. 08.8 & 57700.8 & 594.3 & {\it F814W} & 1266 & 24.21 \pm 0.05 & 37.44 \pm 0.02 \\
  2016 Nov. 08.8 & 57700.8 & 594.3 & Optical     & \nodata & \nodata & 38.46 \pm 0.01 \\
  2016 Dec. 25.6 & 57747.6 & 641.1 & {\it F555W} & 1275 & 24.58 \pm 0.03 & 37.99 \pm 0.01 \\
  2016 Dec. 25.6 & 57747.6 & 641.1 & {\it F814W} & 1275 & 24.81 \pm 0.08 & 37.20 \pm 0.03 \\
  2016 Dec. 25.6 & 57747.6 & 641.1 & Optical     & \nodata & \nodata & 38.30 \pm 0.01 \\
  2017 Feb. 16.2 & 57800.2 & 693.7 & {\it F438W} & 1266 & 25.71 \pm 0.09 & 37.46 \pm 0.04 \\
  2017 Feb. 16.2 & 57800.2 & 693.7 & {\it F555W} & 1266 & 24.98 \pm 0.03 & 37.83 \pm 0.01 \\
  2017 Feb. 16.2 & 57800.2 & 693.7 & {\it F625W} & 1329 & 25.55 \pm 0.07 & 37.30 \pm 0.03 \\
  2017 Feb. 16.2 & 57800.2 & 693.7 & {\it F814W} & 1329 & 25.39 \pm 0.11 & 36.97 \pm 0.04 \\
  2017 Feb. 16.2 & 57800.2 & 693.7 & Optical     & \nodata & \nodata & 38.10 \pm 0.01 \\
  2017 Apr. 09.4 & 57852.4 & 745.9 & {\it F555W} & 1275 & 25.38 \pm 0.05 & 37.67 \pm 0.02 \\
  2017 Apr. 09.4 & 57852.4 & 745.9 & {\it F625W} & 1275 & 25.74 \pm 0.08 & 37.23 \pm 0.03 \\
  2017 Apr. 09.4 & 57852.4 & 745.9 & Optical     & \nodata & \nodata & 37.92 \pm 0.02 \\
  2017 May  04.6 & 57877.6 & 771.1 & {\it F438W} & 1266 & 26.55 \pm 0.18 & 37.13 \pm 0.07 \\
  2017 May  04.6 & 57877.6 & 771.1 & {\it F555W} & 1266 & 25.61 \pm 0.06 & 37.58 \pm 0.02 \\
  2017 May  04.6 & 57877.6 & 771.1 & {\it F625W} & 1329 & 26.28 \pm 0.14 & 37.01 \pm 0.05 \\
  2017 May  04.6 & 57877.6 & 771.1 & {\it F814W} & 1329 & >26.7 & <36.4 \\
  2017 May  04.6 & 57877.6 & 771.1 & Optical     & \nodata & \nodata & 37.82 \pm 0.03 \\
  2017 Jun. 18.2 & 57922.2 & 815.7 & {\it F555W} & 2958 & 25.85 \pm 0.05 & 37.48 \pm 0.02 \\
  2017 Jun. 18.2 & 57922.2 & 815.7 & Optical     & \nodata & \nodata & 37.67 \pm 0.03 \\
  2017 Jun. 28.0 & 57933.0 & 826.5 & {\it F625W} & 2700 & 26.86 \pm 0.14 & 36.78 \pm 0.06 \\
  2017 Jun. 28.0 & 57933.0 & 826.5 & Optical     & \nodata & \nodata & 37.62 \pm 0.03 \\
  2017 Aug. 06.3 & 57971.3 & 864.8 & {\it F438W} & 1266 & >27.4 & <36.8 \\
  2017 Aug. 06.3 & 57971.3 & 864.8 & {\it F555W} & 1395 & 26.68 \pm 0.10 & 37.15 \pm 0.04 \\
  2017 Aug. 06.3 & 57971.3 & 864.8 & {\it F625W} & 1266 & 27.43 \pm 0.30 & 36.55 \pm 0.12 \\
  2017 Aug. 06.3 & 57971.3 & 864.8 & {\it F814W} & 1266 & >26.7 & <36.4 \\
  2017 Aug. 06.3 & 57971.3 & 864.8 & Optical     & \nodata & \nodata & 37.39 \pm 0.05 \\  
  2017 Oct. 02.9 & 58028.9 & 922.4 & {\it F555W} & 2958 & 26.90 \pm 0.10 & 37.06 \pm 0.04 \\
  2017 Oct. 02.9 & 58028.9 & 922.4 & Optical     & \nodata & \nodata & 37.24 \pm 0.04 \\
  2017 Dec. 04.2 & 58091.2 & 984.7 & {\it F555W} & 2790 & 27.51 \pm 0.16 & 36.82 \pm 0.06 \\
  2017 Dec. 04.2 & 58091.2 & 984.7 & Optical     & \nodata & \nodata & 37.01 \pm 0.06 \\
  2018 Jan. 28.0 & 58146.0 & 1039.5 & {\it F555W} & 9069 & 28.46 \pm 0.20 & 36.44 \pm 0.08 \\
  2018 Jan. 28.0 & 58146.0 & 1039.5 & Optical & \nodata & \nodata & 36.69 \pm 0.08 \\
 \enddata
 \textbf{Note.} All photometry is measured using PSF-fitting photometry with {\sc Dolphot}. ``Optical'' refers to the pseudo-bolometric luminosities, in the wavelength range $3500$--$10000$~\AA, derived in Section~\ref{subsec:pbl}. These observations can be found in MAST at \href{http://archive.stsci.edu/doi/resolve/resolve.html?doi=10.17909/T9109B}{DOI:10.17909/T9109B}.
\end{deluxetable}

\begin{deluxetable}{CCC}
 \tablecaption{PSF-fitting $F350LP$ photometry of SN 2012cg. \label{table:mags_12cg}}
 \tablehead{
 \colhead{Phase}  & \colhead{Magnitude}  & \colhead{Luminosity} \\
 \colhead{(days)} & \colhead{(Vega mag)} & \colhead{${\rm log}(L/{\rm erg~s^{-1}})$} 
 }
 \decimals
 \startdata
  567.9  & 23.14 \pm 0.01 & 38.540 \pm 0.003 \\
  575.2  & 23.15 \pm 0.01 & 38.536 \pm 0.004 \\
  582.8  & 23.22 \pm 0.01 & 38.507 \pm 0.004 \\
  591.1  & 23.32 \pm 0.01 & 38.470 \pm 0.004 \\
  597.6  & 23.47 \pm 0.01 & 38.408 \pm 0.004 \\
  602.9  & 23.41 \pm 0.01 & 38.433 \pm 0.004 \\
  609.0  & 23.57 \pm 0.01 & 38.368 \pm 0.004 \\
  616.8  & 23.64 \pm 0.01 & 38.342 \pm 0.005 \\
  625.8  & 24.00 \pm 0.01 & 38.196 \pm 0.006 \\
  632.5  & 24.03 \pm 0.01 & 38.184 \pm 0.006 \\
  642.6  & 23.80 \pm 0.01 & 38.278 \pm 0.006 \\
  924.5  & 25.76 \pm 0.02 & 37.491 \pm 0.009 \\
  976.9  & 25.89 \pm 0.04 & 37.44 \pm 0.015 \\
  1055.6 & 26.32 \pm 0.05 & 37.27 \pm 0.021 \\
 \enddata
 \textbf{Note.} The {\sc Dolphot} {\it wfc3mask} masking routine masked several SN pixels in the {\sc flt} images of $625.8$ and $632.5$ days, leading to depressed flux measurements. These measurements are not used in the various fits described in Section~\ref{sec:decay}.
\end{deluxetable}

To test whether SN 2015F is contaminated by light echoes, we measure its $B-V$ and $V-R$ colors, and compare them to those of SN 2011fe at similar times (as measured by \citealt{2017ApJ...841...48S}). We also compare the colors of both SNe to their colors at maximum light. For SN 2011fe, we use the photometry measured by \citet{2013NewA...20...30M}; for SN 2015F, we use photometry from \citet{2017MNRAS.464.4476C}. All magnitudes have been corrected for both host-galaxy and Galactic extinction. We converted our \textit{HST} photometry in \textit{F438W}, \textit{F555W}, and \textit{F625W} to $B$, $V$, and $R$ by accounting for the shapes of their respective transmission curves.

\begin{figure}
 \centering
 \includegraphics[width=0.47\textwidth]{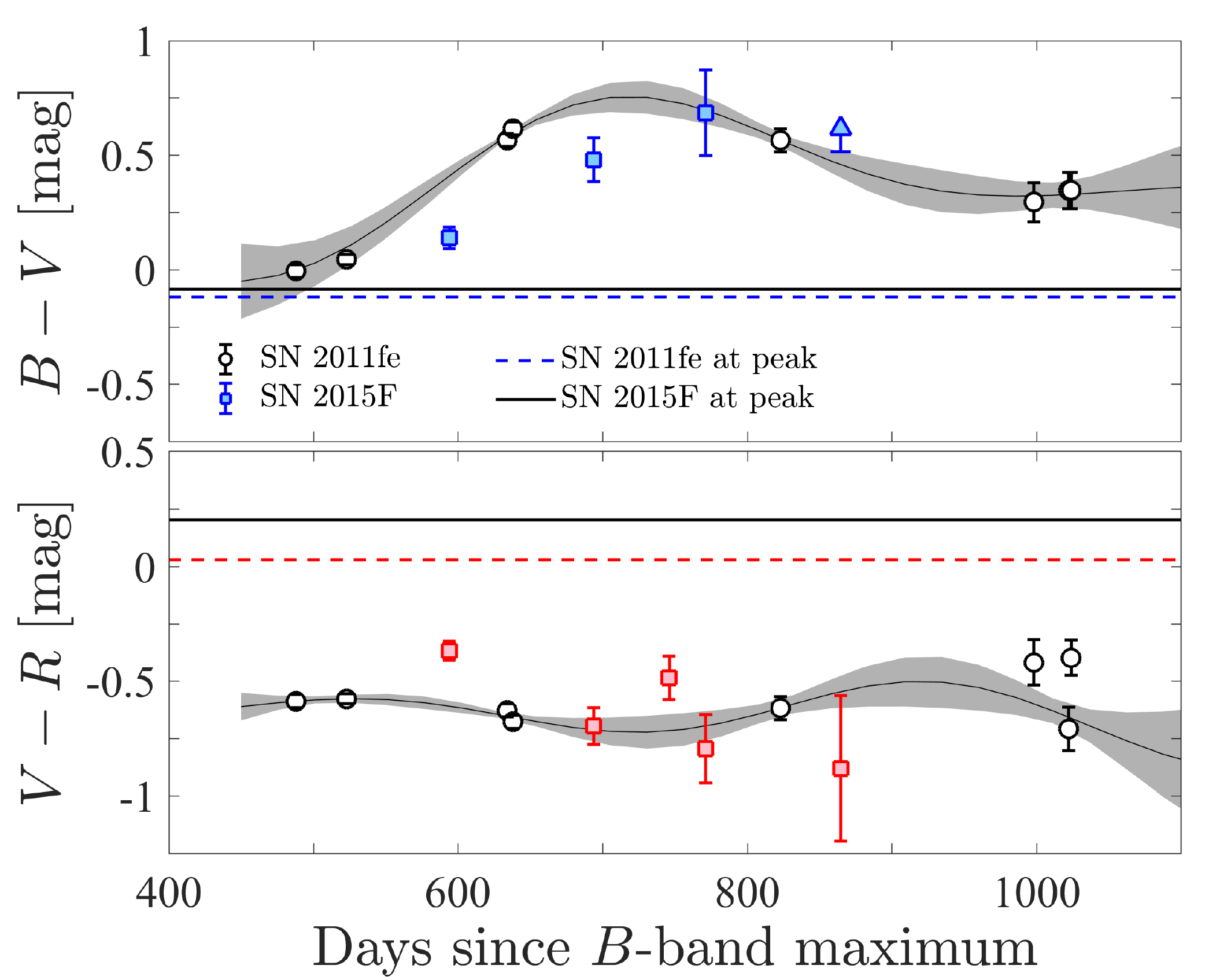}
 \caption{A comparison between the $B-V$ (top) and $V-R$ (bottom) colors of SNe 2015F (blue or red filled squares) and 2011fe (black circles). In each panel, the solid and dashed curves represent the $B-V$ or $V-R$ colors of SNe 2011fe and 2015F, respectively, at $B$-band maximum light. The gray shaded bands connecting the colors of SN 2011fe are Gaussian Process regressions; the width of the band represents the 68\% uncertainty of the fit. The colors of SN 2015F are broadly consistent with those of SN 2011fe. Importantly, both SNe are redder in $B-V$ and bluer in $V-R$ than their colors at peak light. A light echo, which shifts the spectrum bluewards, cannot account for such an effect.}
 \label{fig:color}
\end{figure}

As we show in Figure~\ref{fig:color}, compared to its peak colors, SN 2015F at $600$--$900$ days is redder in $B-V$ by $\sim0.3$--$0.6$ mag, but bluer in $V-R$ by $\sim0.5$--$0.8$ mag. Light echoes preferentially scatter light to bluer wavelengths, making the SNe appear to be bluer than they were at peak. This could explain the blue $V-R$ color. However, the SN should then be even bluer at bluer wavelengths. Instead, the SN is redder in $B-V$. One would have to invoke an exotic source of reddening that would manage to redden the SN in $B-V$ enough to counteract the light echo, yet keep it blue at $V-R$.

Moreover, when SN 2015F is compared to SN 2011fe at the same phase, the two SNe appear to have consistent colors. Late-time spectra of SN 2011fe have shown conclusively that it was not contaminated by light echoes (e.g., \citealt{2015MNRAS.454.1948G}). The similarity of SN 2015F to SN 2011fe, as well as its colors, lead us to conclude that it was not contaminated by a light echo at $600$--$900$ days.

We further argue that, because we do not detect a light echo in SN 2015F at $600$ days, we would not have expected to observe light echo contamination at any of the other phases up to $1040$ days, the range covered in this work. A light echo at later phases will only be possible if the light from the SN encounters a dust sheet within the $\sim440$ days the light continued to travel out between $600$ and $1040$ days. The geometry of the light echo parabola bounds the distance to this dust sheet to $\approx 220$ light days at the shortest (if the light echo would have to travel to the dust sheet and back towards the SN and the observer, along the direct line of sight) and $\approx 660$ light days at the longest. The latter constraint derives from two considerations: the farther the light has to travel until it reaches the dust sheet, the lower the echo's resultant surface brightness and the larger the angle of separation between the SN and the echo. As we do not detect a resolved light echo, this angle cannot be large. Thus, for a light echo to manifest at $1040$ days but not at $600$ days, the dust sheet would have to lie $\approx 1 \pm 0.5$ light years away from the distance reached by the SN light up to $600$ days. We argue that such a small distance is too fine tuned to be probable (though see \citealt{2013ApJ...770L..35S}).

\subsubsection{SN 2012cg}
\label{subsubsec:le12cg}

With the data available at the time, \citet{2016ApJ...819...31G} were unable to rule out the possibility of light echoes contaminating the light curve of SN 2012cg. However, since then, \citet{2016ApJ...820...92M} published the early UVOIR light curve of this SN, which makes it possible to compare the $V-H$ color of the SN at peak to that at late time. As noted above, a SN light curve contaminated by light echoes is expected to have the same colors---or bluer---than it had at peak. \citet{2017ApJ...841...48S} noted that SN 2011fe had $B-H=-0.85 \pm 0.04$ mag at peak, and $F438W-F160W=2.3 \pm 0.1$ mag at 1840 days, making SN 2011fe redder by a factor of $\sim 20$ than it was at peak and placing severe constraints on any contamination of the flux by a light echo. Likewise, SN 2012cg had $V-H = -0.68 \pm 0.07$ mag at $B$-band maximum light \citep{2016ApJ...820...92M} and $F555W-F160W = 2.6 \pm 0.2$ mag at 626 days. Following the same argument, we conclude that SN 2012cg was not contaminated by light echoes either.  
  
\floattable
\begin{deluxetable}{lCCCCCCC}
 \tablecaption{Light curve parameters of SNe Ia observed at late times. \label{table:params}}
 \tablehead{
 \colhead{SN name} & \colhead{SiFTO $s$} & \colhead{$M(^{56}{\rm Co})$\tablenotemark{a}} & \colhead{$M(^{57}{\rm Co})/M(^{56}{\rm Co})$\tablenotemark{b}} & \colhead{$\chi^2/{\rm DOF}$} & \colhead{$t_{\rm freeze_{50}}$\tablenotemark{c}} & \colhead{$\chi^2/{\rm DOF}$} & \colhead{$\Delta L_{900}$\tablenotemark{d}} \\
 \colhead{}        & \colhead{}          & \colhead{$(M_\sun)$}         & \colhead{}                                    & \colhead{}                   & \colhead{(days)}                 & 
 \colhead{}        & \colhead{}
 }
 \decimals
 \startdata
  SN 2011fe                  & 0.969 \pm 0.010 & 0.117^{+0.003}_{-0.004} & 0.043^{+0.004}_{-0.004} & 22/18  &  920^{+20}_{-20} & 25/18  & 0.95 \pm 0.04 \\
  SN 2012cg\tablenotemark{e} & 1.063 \pm 0.011 & 0.156^{+0.007}_{-0.007} & 0.069^{+0.018}_{-0.019} & 2.5/12 &  820^{+60}_{-50} & 2.8/12 & 0.90 \pm 0.08 \\
                             &                 & 0.132^{+0.001}_{-0.001} & 0.072^{+0.002}_{-0.002} & 230/10 &  801^{+7}_{-7}   & 240/10 & 0.89 \pm 0.01 \\ 
  SN 2014J                   & 1.086 \pm 0.010 & 0.140^{+0.014}_{-0.014} & 0.129^{+0.021}_{-0.016} & 0.2/2  &  644^{+40}_{-50} & 0.1/2  & 0.82 \pm 0.03 \\
  SN 2015F                   & 0.906 \pm 0.005 & 0.167^{+0.003}_{-0.003} & 0.004^{+0.003}_{-0.002} & 14/9   & 1300^{+130}_{-80} & 14/9  & 1.13 \pm 0.03 \\
 \enddata
 \tablenotetext{a}{The $^{56}$Co value fit by Equation~\ref{eq:lum}. This represents only a fraction of the total $^{56}$Co produced by the SN, as encapsulated by the light emitted in the range $\approx 3500$--$10000$~\AA.}
 \tablenotetext{b}{The mass ratio of $^{57}$Co and $^{56}$Co, as measured with Equation~\ref{eq:lum}.}
 \tablenotetext{c}{The time at which half of the SN flux is due to freeze-out.}
 \tablenotetext{d}{The difference between the pseudo-bolometric luminosity at $600$ and $900$ days, calculated as $\Delta L_{900} = {\rm log}(L_{600}/L_{900})$.}
 \tablenotetext{e}{The first and second rows show the results when using either the aperture or PSF-fitting photometry of SN 2012cg, respectively.}
\end{deluxetable}

\subsection{Radioactive decay}
\label{subsec:decay}

As in \citet{2016ApJ...819...31G}, we fit the pseudo-bolometric light curve with the solution to the Bateman equation from \citet{2014ApJ...792...10S}:
\begin{equation}\label{eq:lum}
 \begin{split}
  L_A(t) = & 2.221 \frac{B}{A} \frac{\lambda_A}{\rm days^{-1}} \frac{M(A)}{M_\odot} \frac{q_A}{{\rm keV}} {\rm exp}(-\lambda_A t) \\
  & \times10^{43}~{\rm erg~s^{-1}},
 \end{split}
\end{equation}
where $B$ is the fraction of light emitted by the SN in the optical range observed here; $A$ is the atomic number of the decaying nucleus; $\lambda_A$ is the inverse of the half-life time of the decay chain; $q_A$ is the average energy per decay carried by X-rays and charged leptons \citep{2009MNRAS.400..531S}; and $t$ is the time since explosion (which, relative to $B$-band maximum light, occurred at $-17.5 \pm 0.6$ days; \citealt{2015ApJS..221...22I}). 

We fit for the mass of $^{56}$Co, times the bolometric correction $B$, and for the ratio of the mass of $^{57}$Co to $^{56}$Co. Because \citet{2016ApJ...819...31G} showed that the light curves of SNe Ia are not sensitive to the decay of $^{55}$Fe in the phase range probed here, we do not include this isotope in our fit. In addition, we have confirmed that the isotopic ratio of $^{55}$Fe$/^{57}$Co required to significantly influence the late-time light curves up to $t = 1000$ days exceeds any of a wide range of recently-computed 2D single-degenerate hydrodynamic models by \citet{2017ApJ...841...58D} by more than a factor of 2. Thus, we may safely exclude the contribution of $^{55}$Fe in this work.


For SN 2015F, this decay model results in $M(^{56}{\rm Co})=0.167 \pm 0.003~M_\sun$ and $M(^{57}{\rm Co})/M(^{56}{\rm Co})=0.004^{+0.003}_{-0.002}$, with $\chi^2/{\rm DOF}=14/9$. We note that the $97$\% uncertainty of the mass ratio makes the measurement consistent with zero.

To compare the $M(^{56}{\rm Co})$ value measured from light emitted in the wavelength range $\sim3500$--$10000$~\AA\ to the total $M(^{56}{\rm Co})$ of SN 2015F, we first estimate the latter by fitting a straight line to the $M(^{56}{\rm Ni})$ measured by \citet{2015MNRAS.454.3816C} to their SiFTO stretch values \citep{2008ApJ...681..482C}. We estimate a total $M(^{56}{\rm Ni})$ of $0.40 \pm 0.05~M_\sun$, which implies that $0.42 \pm 0.04$ of the light from SN 2015F is emitted in the wavelength range $\sim3500$--$10000$~\AA. This last value is consistent with the bolometric correction measured by \citet{2017ApJ...841...48S} for SN 2011fe in the the same wavelength range.

The derived $^{57}$Co$/^{56}$Co mass ratio is seven times smaller than the same ratio measured in SN 2012cg. To make a similar comparison with the ratios measured for SNe 2011fe and 2014J, we fit the model in Equation~\ref{eq:lum} to their pseudo-bolometric luminosities, as measured by \citet{2017MNRAS.468.3798D} and \citet{2018ApJ...852...89Y}, respectively. For all of the SNe, we fit the data starting at 500 days after maximum light, and use their individual rise times: $\approx18.6$ days for SNe 2011fe and 2014J \citep{Nugent2011,2015ApJ...799..105S}, $\approx 17.3$ days for SN 2012cg \citep{2012ApJ...756L...7S}, and $\approx17.5$ days for SN 2015F \citep{2015ApJS..221...22I}. Varying the rise time used in Equation~\ref{eq:lum} has a negligible impact on the value of the $^{57}$Co$/^{56}$Co mass ratio. For all four SNe, the pseudo-bolometric light curve was constructed in roughly the same wavelength range, i.e., $\approx 3500$--$10000$~\AA. The results from these fits are tabulated in Table~\ref{table:params}.

\begin{figure}
 \centering
 \begin{tabular}{c}
  \includegraphics[width=0.47\textwidth]{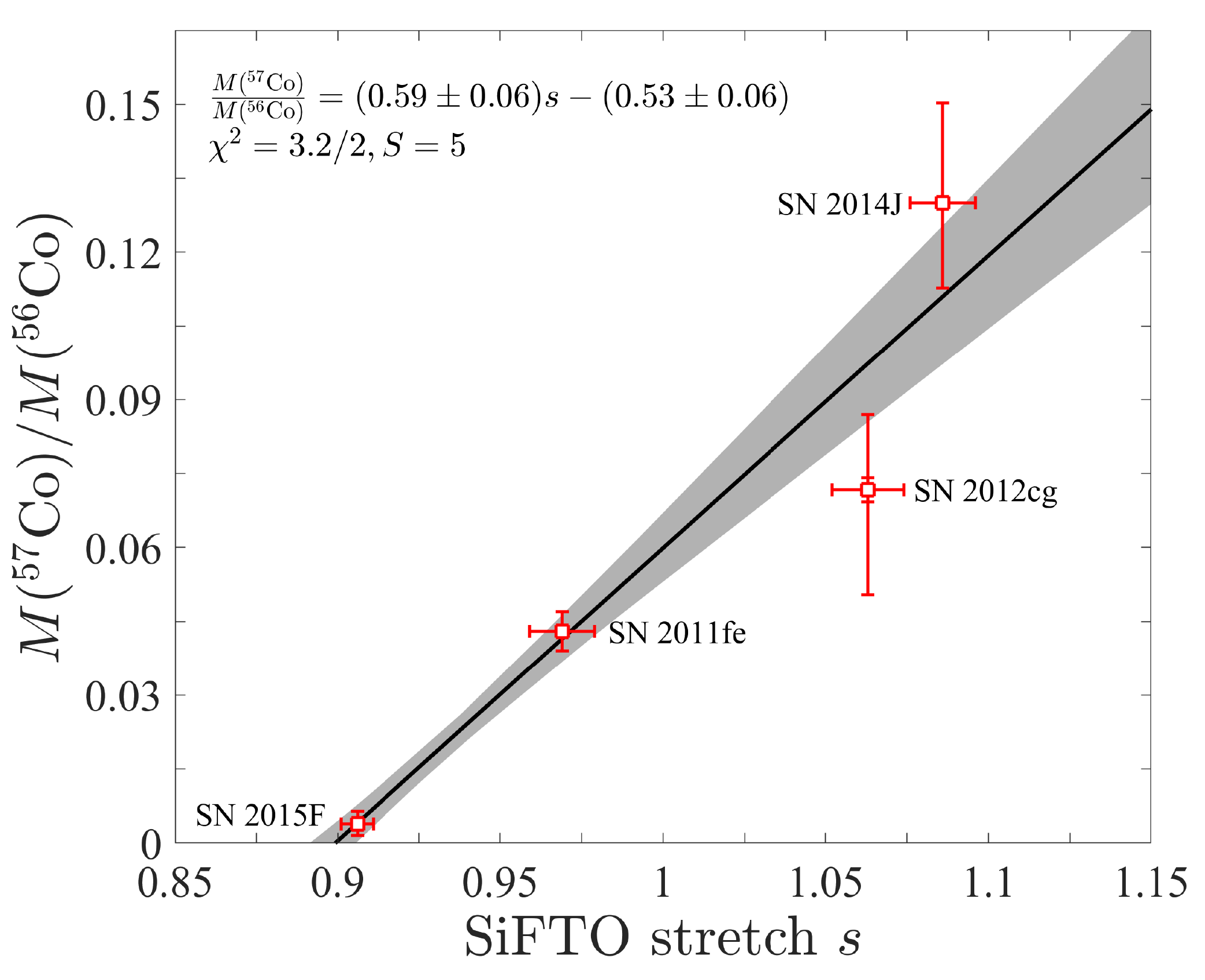} \\ 
  \includegraphics[width=0.47\textwidth]{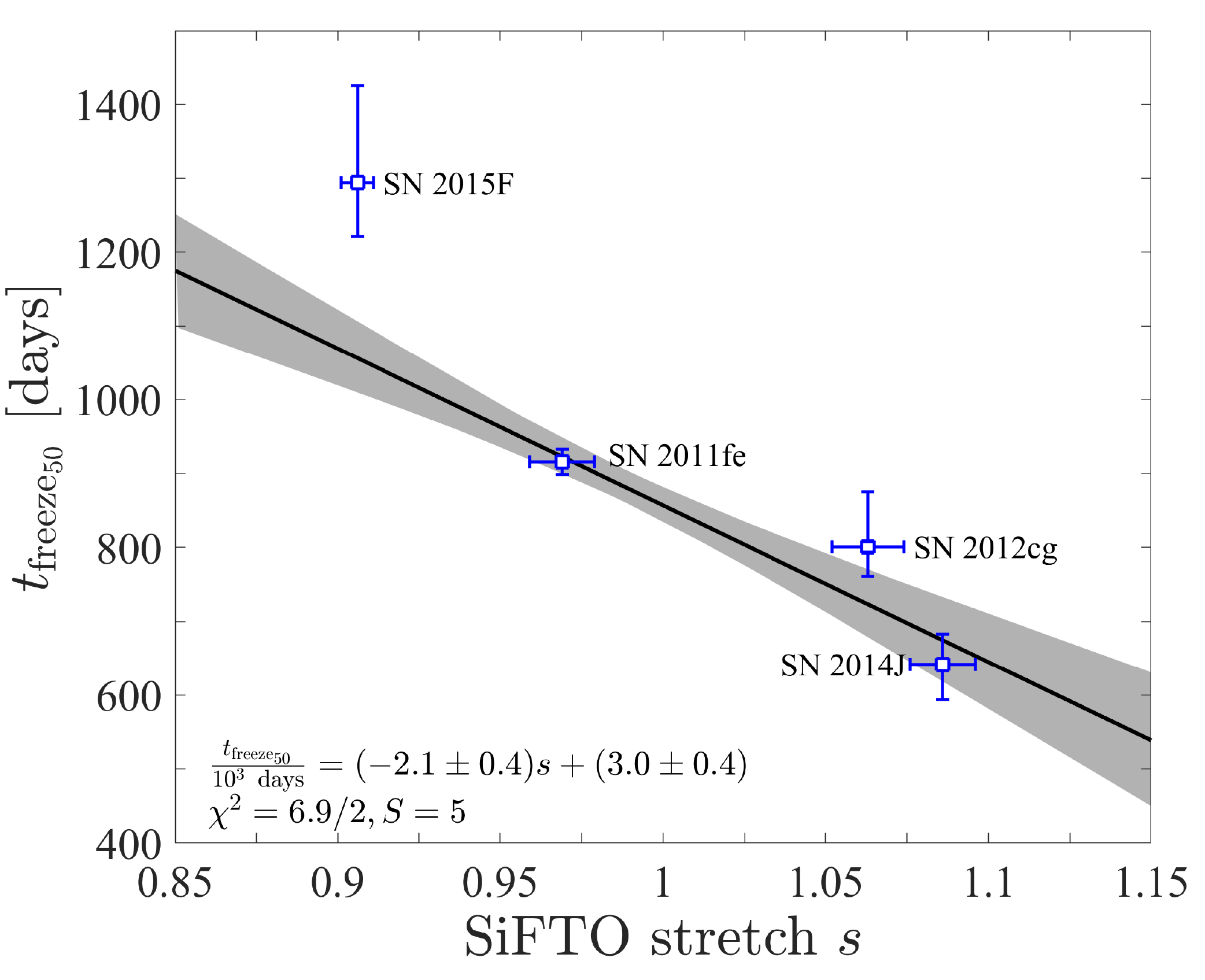} \\
 \end{tabular} 
 \caption{The mass ratio of $^{57}$Co and $^{56}$Co (top panel) or $t_{\rm freeze_{50}}$, the time at which freeze-out dominates the light curve (bottom panel), vs.\ the stretch of SNe 2011fe, 2012cg, 2014J, and 2015F (squares). In both panels, a likelihood ratio test finds that a linear fit (solid black curve) is preferred over a constant fit with a significance of $S>5\sigma$, indicating a possible correlation between the stretch of an SN Ia (a direct proxy for its luminosity) and the $^{57}$Co$/^{56}$Co ratio produced in the explosion or the time at which freeze-out occurs. The gray region represents the 68\% uncertainty of the linear fit. SN 2012cg has a double error bar, representing the different uncertainties afforded by aperture and PSF-fitting photometry. Contamination by light echoes has been ruled out for all four SNe Ia.}
 \label{fig:corr_model}
\end{figure}

In the top panel of Figure~\ref{fig:corr_model}, we plot the $^{57}$Co$/^{56}$Co mass ratios as a function of the SiFTO stretch parameter, $s$, where SNe with larger $s$ values are intrinsically more luminous than those with smaller values. 

The data have a Pearson's correlation coefficient of $\rho=0.93$ with a nominal $p$-value of $0.07$. To account for the measurement uncertainties, we run $1000$ trials in which we randomly vary both the stretches and $^{57}$Co$/^{56}$Co mass ratios according to their uncertainties (and assuming those uncertainties are Gaussian). When we use either the aperture or PSF-fitting photometry of SN 2012cg, the correlation is deemed significant ($p$-value $\leqslant 0.05$) in $38$\% or $36$\% of the trials, respectively.

A better way to account for the measurement uncertainties is to use the likelihood ratio test (see \citealt{2017ApJ...837..120G}), which we use to test whether a 1st-order polynomial (a linear fit) is preferred over a 0th-order polynomial (a constant). According to this test, there is a strong preference (with a significance of $S>5\sigma$) for a linear fit of the form $M(^{57}{\rm Co})/M(^{56}{\rm Co})=(0.59\pm0.06)s-(0.53\pm0.06)$ with $\chi^2/{\rm DOF}=3.2/2$. When using the more precise PSF-fitting photometry of SN 2012cg, the best-fitting linear function is $M(^{57}{\rm Co})/M(^{56}{\rm Co})=(0.500\pm0.004)s-(0.450\pm0.002)$ with $\chi^2/{\rm DOF}=22/2$. The higher precision of the SN 2012cg measurement pulls the fit down and causes SN 2014J to appear as an outlier. The significance of the likelihood ratio test does not depend on the type of photometry used for SN 2012cg.

Taken together, the statistical tests above imply a possible correlation between the derived ratio of $M(^{57}{\rm Co})/M(^{56}{\rm Co})$ and $s$, with more luminous SNe Ia producing a higher ratio of these two Co isotopes. However, because our analysis is based on a sample of only four SNe, a larger sample is required to thoroughly test the claim of a correlation.

The assumptions that went into the construction of the pseudo-bolometric light curve in Section~\ref{subsec:pbl} provide a possible source of systematic uncertainty in the derivation of the $^{57}$Co$/^{56}$Co mass ratio of SN 2015F. We assume that throughout the phase range probed here, the color evolution of SN 2015F remains flat and most of the light comes out in $F555W$. \citet{2017ApJ...841...48S} show that this is indeed the case for SN 2011fe, but also that at later times, $>1100$ days after explosion, the spectrum of SN 2011fe moves from being dominated by light coming out in $F555W$ to that coming out in $F438W$. Because our $F438W$ observations do not extend past $\sim 860$ days, it is possible that we missed a rebrightening of the SN in that band before our final visit. To test this possibility, we replace our $F438W$ observations of SN 2015F with the $B$-band observations of SN 2011fe taken by \citet{2017ApJ...841...48S}, normalized to the observed $F438W$ magnitude of SN 2015F at 600 days. We then reconstruct the pseudo-bolometric light curve of SN 2015F and fit it with the radioactive decay model in Equation~\ref{eq:lum}. The resultant $^{57}$Co$/^{56}$Co mass ratio is $0.020 \pm 0.002$, five times higher than our initial result, but still half the mass ratio measured for SN 2011fe. If, on top of the $B$-band data, we add the $R$-band data of SN 2011fe as well, the mass ratio rises by an additional 25\% to $0.025 \pm 0.002$. \citet{2017ApJ...841...48S} did not observe SN 2011fe in $I$, but based on the spectral evolution of SN 2011fe, its inclusion should have a similar, or smaller, effect as the $R$-band data. Thus, although it is possible that our lack of multicolor observations at the end of our program have led us to underestimate the $^{57}$Co$/^{56}$Co mass ratio of SN 2015F, using SN 2011fe observations still results in a ratio that is roughly half that of SN 2011fe, consistent with the correlation shown here.

\subsection{Freeze-out}
\label{subsec:freeze}

As in \citet{2017MNRAS.472.2534K}, we assume that the amount of light produced by the freeze-out effect depends on the electron and ion densities of the ejecta, $n_e$ and $n_{\rm ion}$, respectively, and its volume, $V$, as $n_e \times n_{\rm ion} \times V \propto t^{-3}$. We fit all four SNe with a model that combines this effect with $^{56}$Co and measure $t_{\rm freeze_{50}}$, which we define as the time at which freeze-out contributes half of the total pseudo-bolometric flux. The results of these fits are shown in Table~\ref{table:params}.

In the bottom panel of Figure~\ref{fig:corr_model}, we show that $t_{\rm freeze_{50}}$ appears to be anti-correlated with stretch. The data have a Pearson's correlation coefficient of $\rho=-0.95$, with a $p$-value of $0.05$. In $1000$ randomized trials, the correlation is found to be significant $33$\% and $42$\% of the time when either the aperture or PSF-fitting photometry of SN 2012cg is used.

A likelihood ratio test results in a significance of $S>5\sigma$, whether we use the aperture or PSF-fitting photometry of SN 2012cg. We express this correlation as $t_{\rm freeze_{50}} / (10^3~{\rm days}) = (-2.1 \pm 0.4)s +(3.0 \pm 0.4)$, with $\chi^2/{\rm DOF}=6.9/2$ (aperture photometry) or $t_{\rm freeze_{50}} /(10^3~{\rm days}) = (-1.4 \pm 0.2)s +(2.3 \pm 0.2)$, with $\chi^2/{\rm DOF}=12/2$ (PSF-fitting photometry).

\subsection{A model-independent light-curve characterization}
\label{subsec:nonparam}

The late-time boost to SN Ia light curves could also come from an evolving positron trapping fraction. \citet{2017MNRAS.472.2534K} and \citet{2017MNRAS.468.3798D} showed that models with varying positron fractions provided fits consistent with the observations of SN 2011fe. However, because taking this effect into account is not straightforward, we do not model it here. Instead, we prefer to offer a model-independent description of the possible correlation between the stretch of the SNe Ia and the shape of their late-time light curves. This purely observational correlation can then be used to test not only the heating models described here, but any models suggested in the future as well.

We draw on the Phillips width-luminosity relation, which is parametrized by either $\Delta m_{15}(B)$ or stretch, and offer the following recipe:
\begin{enumerate}
 \item Follow a SN Ia out to at least 900 days past maximum light.
 \item Use colors to rule out light-echo contamination.
 \item Construct a pseudo-bolometric light curve in the optical wavelength range, $\approx 3500$--$10000$~\AA.
 \item Measure the ratio between the pseudo-bolometric luminosities at $600$ and $900$ days, $\Delta L_{900} = {\rm log}(L_{600}/L_{900})$.
\end{enumerate}
At $600$ days, the light curves of all SNe Ia are still supposed to follow the radioactive decay of $^{56}$Co; at $900$ days, some deviation from this state should already have occured, at least according to the explosion models suggested so far. Though it would be more informative to probe the SNe at later phases, that is only possible for the very few SNe discovered at $<10$~Mpc. 

To measure $L_{600}$ and $L_{900}$ we fit straight lines to the pseudo-bolometric luminosities of the SNe in the phase ranges $500<t<800$ days and $800<t<1200$ days, then sample the fits at $600$ and $900$ days. In the case of SN 2014J, which only has four measurements at $t>500$ days, we linearly interpolate the luminosities and their uncertainties at these phases. In the case of SN 2012cg, a linear fit to the three $t>900$ photometry points of this SN (at $t=925$, $977$, and $1056$ days) produces a $\Delta L_{900}$ uncertainty twice as large as the one that results from linearly extrapolating the value at $900$ days. The linear fit thus overestimates the uncertainty on $\Delta L_{900}$, as a photometry measurement at $900$ days would have a smaller, not higher, uncertainty than that of the nearest measurement at $925$ days. Both techniques result in the same $\Delta L_{900}$ value, and we choose to report the smaller of the two uncertainties. The $\Delta L_{900}$ values for all four SNe Ia are reported in Table~\ref{table:params}. 

\begin{figure}
 \centering
 \includegraphics[width=0.47\textwidth]{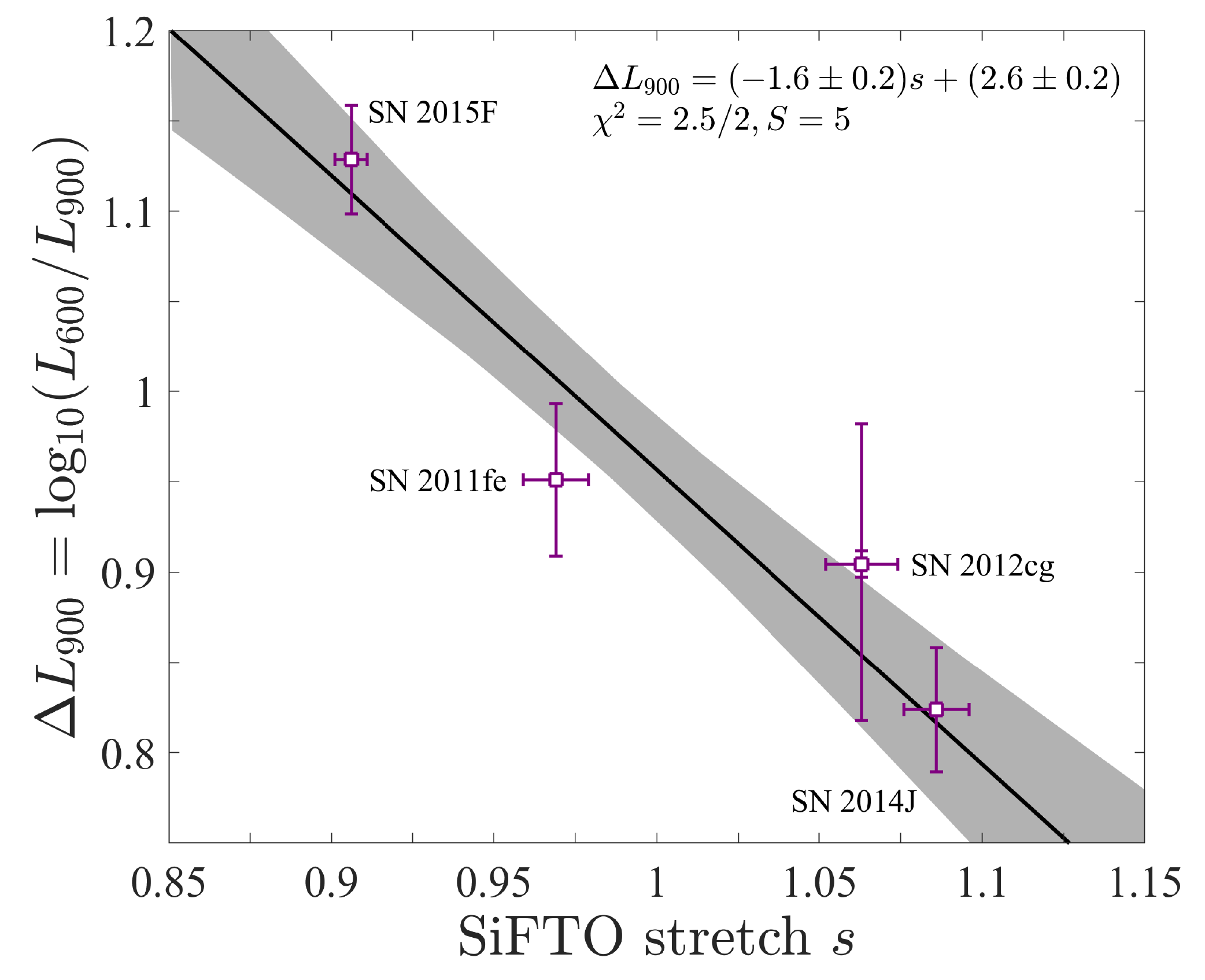}
 \caption{$\Delta L_{900}$ vs.\ the stretch of SNe 2011fe, 2012cg, 2014J, and 2015F (purple squares). A likelihood ratio test finds that a linear fit (solid black curve) is preferred over a constant fit with a significance of $S>5\sigma$, indicating a possible correlation between the stretch of an SN Ia (a direct proxy for its luminosity) and the shape of its light curve at $>900$ days after maximum light. The gray region represents the 68\% uncertainty of the linear fit. SN 2012cg has a double error bar to show the uncertainties resulting from the use of aperture and PSF-fitting photometry.}
 \label{fig:corr_non}
\end{figure}

In Figure~\ref{fig:corr_non}, we plot $\Delta L_{900}$ vs.\ stretch. These data have a Pearson's $\rho=-0.94$ with a nominal $p$-value of $0.06$. In $1000$ random trials, the correlation is significant $31$\% and $40$\% of the time, when using either the aperture or PSF-fitting photometry of SN 2012cg, respectively.

The likelihood ratio test shows that a linear fit is preferred over a constant at $S>5\sigma$ when using either the aperture or PSF-fitting photometry of SN 2012cg. The shape of the linear fit does not vary significantly when using either set of measurements: the less precise aperture photometry results in $\Delta L_{900} = (-1.6 \pm 0.2)s + (2.6 \pm 0.2)$, with $\chi^2/{\rm DOF}=2.5/2$, while the more precise PSF-fitting photometry produces $\Delta L_{900} = (-1.3 \pm 0.2)s + (2.3 \pm 0.2)$, with $\chi^2/{\rm DOF}=5.6/2$. In Figure~\ref{fig:corr_non}, we show the first of these two fits.

\section{Conclusions}
\label{sec:discuss}

We have used \hst\ to observe SN 2015F in the wavelength range $\sim3500$--$10000$~\AA\ when the SN was $\approx 600$--$1040$ days past maximum light. As opposed to SNe 2011fe, 2012cg, and 2014J, whose pseudo-bolometric light curves begin to flatten out in this phase range, the light curve of SN 2015F remains consistent with being powered solely by the radioactive decay of $^{56}$Co. 

Instead of thinking of SN 2015F as significantly different than the three other SNe Ia mentioned above, we show that the late-time light curve of each of these SNe is distinct. When all four SNe Ia are fit with a model that assumes the combined radioactive decays of $^{56}$Co and $^{57}$Co, as suggested by \citet{2009MNRAS.400..531S}, there appears to be a correlation between the estimated Co mass ratios and the stretch of the SNe. If, instead, the late-time boost to the light curve is due to freeze-out, as suggested by \citet{1993ApJ...408L..25F}, then an anti-correlation appears between stretch and the time at which freeze-out begins to dominate the light curve.

On their own, each of these correlations can be used to constrain these physical processes. For example, the correlation between $M(^{57}{\rm Co})/M(^{56}{\rm Co})$ and stretch extrapolates to zero for SNe Ia with $s<0.9$. This could mean that subluminous SNe Ia should, as a class, under-produce $^{57}$Co, if the $^{57}$Co$/^{56}$Co ratio is the dominant, physical reason behind the observed correlation. This, in turn, constrains possible explosion models. The freeze-out effect can likewise be constrained with the correlation measured here.

Yet, it is plausible that more than one heating mechanism could be at play. Thus, we suggest a model-independent correlation between the intrinsic luminosity of an SN Ia (encapsulated by its stretch, or $\Delta m_{15}(B)$) and the shape of its late-time light curve. We parametrize the latter as $\Delta L_{900}={\rm log}(L_{600}/L_{900})$, i.e., the difference between the optical luminosity of the SN at $600$ days, when all the light curves are still dominated by the radioactive decay of $^{56}$Co, and $900$ days, when the additional extra heating mechanism is already in play. We measure this correlation to be: $\Delta L_{900} = (-1.6 \pm 0.2)s + (2.6 \pm 0.2)$.

The correlations measured here are based on only four objects; more SNe Ia need to be followed to $>900$ days in order to test and refine the claims made here. Moreover, the four SNe Ia used in this work are all classified as ``normal'' SNe. It remains to be seen whether different subtypes of SNe Ia would fall off these correlations, as is the case with subluminous SNe and the Phillips relation (e.g., \citealt{1996AJ....112.2391H,Phillips1999,2004ApJ...613.1120G,2008MNRAS.385...75T}). 

The in-depth study of the late-time light curves of SNe Ia is quickly maturing and promises to become a strong diagnostic of SN Ia physics. A future paper will present late-time observations of a fifth SN Ia (ASASSN-14lp), and a forthcoming paper by Fisher et al. will examine the correlations measured here in relation to several SN Ia progenitor and explosion scenarios. 


\section*{Acknowledgments}

We thank Linda Dressel, Weston Eck, Mario Gennaro, and Blair Porterfield at the Space Telescope Science Institute for shepherding Programs GO--14611 and 15415, and Daniel Eisenstein, Peter Garnavich, and the anonymous referee for helpful discussions and comments. OG was supported by NASA through \hst-GO--14611 and 15415. IRS acknowledges funding from the Australian Research Council under grant FT160100028. BS was partially supported by NASA through \hst-GO--14166 and 14678 and Hubble Fellowship grant HF-51348.001. This work is based on data obtained with the NASA/ESA {\it Hubble Space Telescope}, all of which was obtained from the Mikulski Archive for Space Telescopes (MAST). Support for Programs GO--14166, GO--14611, GO--14678, and GO--15414 was provided by NASA through grants from the Space Telescope Science Institute, which is operated by the Association of Universities for Research in Astronomy, Incorporated, under NASA contract NAS5-26555. Support for MAST for non-\hst\ data is provided by the NASA Office of Space Science via grant NNX09AF08G and by other grants and contracts. This work utilized the Extreme Science and Engineering discovery Environment (XSEDE), which is supported by National Science Foundation grant number ACI-1053575. Simulations at UMass Dartmouth were performed on a computer cluster supported by NSF grant CNS-0959382 and AFOSR DURIP grant FA9550-10-1-0354. This research has made use of NASA's Astrophysics Data System and the NASA/IPAC Extragalactic Database (NED) which is operated by the Jet Propulsion Laboratory, California Institute of Technology, under contract with NASA. 


\software{Dolphot \citep{2000PASP..112.1383D}, Matlab}


\end{document}